\documentclass[12pt] {article}
\usepackage{amssymb}
\usepackage{amsmath}
\usepackage{amsfonts}
\usepackage{epsfig}
\usepackage{graphics}

\def\A{{\cal A}}
\def\E{{\cal E}}

\def\Iden{\mbox{$\bf 1\ $}}

\def\M{{\cal M}}

\begin{document}
         
\author{ Navin Khaneja \thanks{To whom correspondence may be addressed. Email:navinkhaneja@gmail.com} \thanks{IIT Bombay, Powai - 400076, India.}}

\vskip 4em

\title{\bf Critique of Feynman Propagator, the $\E \cdot x$ gauge.}

\maketitle
  
\vskip 3cm

\begin{center} {\bf Abstract} \end{center}
Consider M\o{}ller scattering. Electrons with momentum $p$ and $-p$ scatter by exchange of photon say in $z$ direction to $p+q$ and $-(p+q)$.
The scattering amplitude is well known, given as Feynman propagator $ \M = \frac{(e \hbar c)^2}{\epsilon_0 V} \frac{\bar{u}(p+q) \gamma^{\mu} u(p) \ \bar{u}(-(p+q)) \gamma_{\mu} u(-p)}{q^2}$, where $V$ is the volume of the scattering electrons, $e$ elementary charge and $\epsilon_0$
permitivity of vacuum. But this is not completely correct.
Since we exchange photon momentum in $z$ direction, we have two photon polarization $x,y$ and hence the true scattering amplitude should be
$$ \M_1 = \frac{(e \hbar c)^2}{\epsilon_0 V} \frac{ \bar{u}(p+q) \gamma^{x} u(p) \ \bar{u}(-(p+q)) \gamma_{x} u(-p)\ \ + \bar{u}(p+q) \gamma^{y} u(p) \ \bar{u}(-(p+q)) \gamma_{y} u(-p) \ }{q^2}. $$ But when electrons are non-relativistic, $\M_1 \sim 0$. This is disturbing, how will we ever get the coulomb potential, where $\M \sim \frac{(e \hbar c)^2}{\epsilon_0 V q^2}$. Where is the problem ? The problem is with the gauge in Dirac equation.

For a plane wave along $z$ direction, with electric field $E_x \sin (kz - \omega t)$, the Lorentz gauge is  $$ (A_0, A_x, A_y, A_z) = \frac{E_x}{\omega} \cos(kz-\omega t)(0, 1, 0, 0)$$. But this gauge is not suited for calculating optical transitions, because we don't recover the Rabi frequency $q E_x d$ ($d$ electric dipole moment). What we find is something orders of magnitude smaller. Nor is it suitable for calculating electron electron scattering because we don't recover Coulomb potential. What we find is something orders of magnitude smaller. Instead, we work with $\E \cdot x$ gauge  $$ (A_0, A_x, A_y, A_z) = \frac{-E_x}{2} ( x\ \sin(kz-\omega t), -\frac{\cos(kz-\omega t)}{\omega}, 0, \frac{x}{c} \sin(kz-\omega t) ) $$ ($c$ light velocity) to find everything correct. What we get is new propagator.

\section{Introduction}

\begin{figure}[htb!]
  \centering
  \includegraphics[scale = .5]{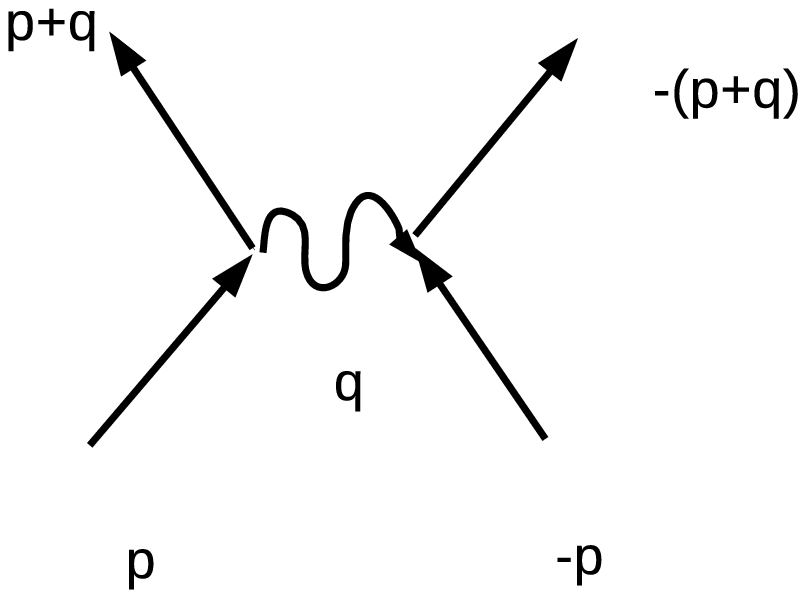}
  \caption{Fig. depicts m\o{}ller scattering. Two electrons with momentum $p$ and $-p$ scatter by exchange of photon to $p+q$ and $-(p+q)$.} \label{fig:moller}
\end{figure}

The heart of interactions in high energy physics is the beautiful electron electron scattering of M\o{}ller. The coulomb interaction between electrons. Fig. \ref{fig:moller} shows two electrons with momentum $p$ and $-p$ scatter by exchange of photon say in $z$ direction to $p+q$ and $-(p+q)$.
The scattering amplitude is well known, given as Feynman propagator $ \M = \frac{(e \hbar c)^2}{\epsilon_0 V} \frac{\bar{u}(p+q) \gamma^{\mu} u(p) \ \bar{u}(-(p+q)) \gamma_{\mu} u(-p)}{q^2}$, \cite{peskin, griffiths.0, thomson}, where $V$ is the volume of the scattering electrons, $e$ elementary charge and $\epsilon_0$
permitivity of vacuum. But this is not completely correct.
Since we exchange photon momentum in $z$ direction, we have two photon polarization $x,y$ and hence the true scattering amplitude should be
$$ \M_1 = \frac{(e \hbar c)^2}{\epsilon_0 V} \frac{ \bar{u}(p+q) \gamma^{x} u(p) \ \bar{u}(-(p+q)) \gamma_{x} u(-p)\ \ + \bar{u}(p+q) \gamma^{y} u(p) \ \bar{u}(-(p+q)) \gamma_{y} u(-p) \ }{q^2}. $$ But when electrons are non-relativistic, $\M_1 \sim 0$. This is disturbing, how will we ever get the coulomb potential, where $\M \sim \frac{(e \hbar c)^2}{\epsilon_0 V q^2}$. Where is the problem ? The problem is with the Dirac equation, it is not all correct in presence of electromagnetic field.

For a plane wave along $z$ direction, with electric field $E_x \sin (kz - \omega t)$, the Lorentz gauge is  $(A_0, A_x, A_y, A_z) = \frac{E_x}{\omega} \cos(kz-\omega t)(0, 1, 0, 0)$. But this gauge is not suited for calculating optical transitions, because we don't recover the Rabi frequency $q E_x d$ ($d$ electric dipole moment). What we find is something orders of magnitude smaller. Nor is it suitable for calculating electron electron scattering because we don't recover Coulomb potential. What we find is something orders of magnitude smaller. Instead, we work with $\E \cdot x$ gauge  $$ (A_0, A_x, A_y, A_z) = \frac{-E_x}{2} ( x\ \sin(kz-\omega t), -\frac{\cos(kz-\omega t)}{\omega}, 0, \frac{x}{c} \sin(kz-\omega t) )$$ ($c$ light velocity) to find everything correct. What we get is a new Feynman propagator.

The paper is organized as follows. We first derive the amplitude of optical transition using Dirac equation with Lorentz gauge. We show it is way too small. We introduce the new gauge, we call the $E \cdot x$ gauge. We show we can correctly calculate the optical Rabi frequency using this gauge. We then use this gauge to calculate M\o{}ller scattering amplitude and show we recover the Coulomb potential.  The result is we get a new Feynman propagator.

\section{Dirac Equation and Lorentz gauge}

\section{Introduction}

Take a classical electron, with coordinates $(x, y, z) = (x_1, x_2, x_3)$ . Its Lagrangian in the electromagnetic field is

\begin{equation}
L = \frac{m}{2} \sum_i \dot{x_i}^2  + q \sum_i A_i \dot{x_i} - q A_0,    
\end{equation} where $q$ and $m$ are electron charge and mass. $\A$ and $V$ are vector and scalar potentials. 

The Euler Lagrange equations are the familiar Lorentz force law $ m \dot {v} = q (E + v \times B)$, where $v$ is the velocity vector, $E_i = - \frac{\partial A_i}{\partial t} - \frac{\partial V}{\partial x_i}$, $B_i = \frac{\partial A_k}{\partial x_j} - \frac{\partial A_j}{\partial x_k}$, the electric and magnetic fields.

The momentum $p_i = \frac{\partial L}{\partial x_i}$ and the Hamiltonian of the system $H = p_i \frac{\partial}{\partial \dot{x}_i} - L$ is

\begin{equation}
H = \sum_{j=x, y, z}  \frac{(p_j - q A_j)^2}{2m} + q A_0.    
\end{equation}  

\subsection{Dirac and Schr\"odinger Equation}

The Electron  Schr\"odinger Equation \cite{Griffiths} is

\begin{equation}
i  \frac{\partial \psi}{\partial t} =   \left ( \sum_{j=x, y, z}  \frac{(-i \hbar \frac{\partial}{\partial x_j} - q A_j)^2}{2m} + q A_0 \right ) \psi,
\end{equation} where $\psi$ is electron wave-function. This equation is not very tractable, because it is nonlinear in $\A$, lets write a linear equation, which is the Dirac equation \cite{thomson}, which takes the form

\begin{equation}
i  \frac{\partial \phi}{\partial t} =   \left ( \sum_{j=x, y, z}  c (-i \ \hbar \frac{\partial}{\partial x_j} - q A_j) \alpha_j  + \beta mc^2 + q A_0 \right ) \phi.
\end{equation} where $\alpha_j = \sigma_z \otimes \sigma_j$ and $\beta = \sigma_x \otimes \Iden$ are Dirac matrices, where $\sigma_j$ are  the Pauli matrices, $\sigma_z = \left ( \begin{array}{cc} 1 & 0 \\ 0 & -1 \end{array} \right )$. $\phi$ is electron spinor, for a electron wave with momentum $k$, takes the form $\phi = \left [ \begin{array}{c} \cos \frac{\theta}{2} \\ \sin \frac{\theta}{2} \end{array} \right ] \otimes {\bf \uparrow}$ , where ${\bf \uparrow}$ is spin up,  $\cos \theta = \frac{\hbar k}{mc} = \frac{\upsilon}{c}$, where $\upsilon = \frac{\hbar k}{m}$, is electron wave group velocity. Electron Orbitals are of size $\sim A^{\circ}$, their $k \sim 10^{10} m$ , then $\upsilon \sim 10^6 m/s$ and $\cos \theta \sim \frac{10^6}{3 \times 10^{8}} \sim 10^{-3}$. Electron is non-relativistic, $\cos \theta = \frac{\upsilon}{c} \sim 0$, $\theta \sim \frac{\pi}{2}$,
$\phi = \frac{1}{\sqrt{2}} \left [ \begin{array}{c} 1  \\ 1 \end{array} \right ] \otimes {\bf \uparrow}$.

To fix ideas, take incoming EM wave,  along $z$ direction, with electric field $E_x \sin (kz - \omega t)$, the Lorentz gauge is  $(A_0, A_x, A_y, A_z) = \frac{E_x}{\omega} \cos(kz-\omega t)(0, 1, 0, 0)$. Electron wave with momentum $q$ absorbs the photon with momentum $k$, and transits to momentum $q + k$. The transition is driven by Dirac matrix $\alpha_x$, with transition amplitude

\begin{equation}
\M =  \left [ \begin{array}{c} \cos \frac{\theta}{2} \\ \sin \frac{\theta}{2} \end{array} \right ] \otimes {\bf \uparrow} (\underbrace{\sigma_z \otimes \sigma_x}_{\alpha_x})  \left [ \begin{array}{c} \cos \frac{\theta}{2} \\ \sin \frac{\theta}{2} \end{array} \right ] \otimes {\bf \downarrow} = q c A_x \frac{\upsilon}{c} = q E_x  \frac{\upsilon}{\omega}  
\end{equation}

If we have electron orbital $\phi_0$ then $k' = \frac{M}{M+m}k$ of photon momentum goes to electron-nuclear relative coordinate, while $k'' = k$ momentum to CM (center of mass), where $M$ is nucleus mass. The process drives the transition

$$ \phi_0 {\bf \uparrow} \ \longrightarrow  \ \exp(i k'z)  \phi_0 {\bf \downarrow}, $$ with amplitude $\M =  q E_x  \frac{\upsilon}{\omega}  $. 

When orbital $\phi_1$ is different from $\phi_0$ we go to $$ \phi_0 {\bf \uparrow} \ \longrightarrow  \ \exp(i k'z)  \phi_0 {\bf \downarrow}, $$ with amplitude $\M =  q c A_x  \frac{\upsilon}{c}  $ whose overlap with $\phi_1$ is

$$ \M_1 = q  c A_x  \frac{\upsilon}{c} i k' \underbrace{\langle \phi_1 |z| \phi_0 \rangle}_{d_z}  = i q E_x d_z \ \frac{\upsilon}{c}, $$ where $c k' \sim \omega$. 

But this is not suited for study of optical transitions, because we don't recover the Rabi frequency $q E_x d$. What we find is orders of magnitude smaller (down by $\frac{\upsilon}{c}$). Instead we work with gauge

$$ (A_0, A_x, A_y, A_z) = \frac{-E_x}{2} ( x\ \sin(kz-\omega t), -\frac{\cos(kz-\omega t)}{\omega}, 0, \frac{x}{c} \sin(kz-\omega t) ). $$

Now we have process driven by $x$ term. For the $\E$ process, the amplitude of $\phi_0 \rightarrow \phi_0$ is just $0$, as $\langle \phi_0 |x| \phi_0 \rangle = 0$ and the  amplitude of $\phi_0 \rightarrow \phi_1$ is simply  

$$ \M_1' = q  E_x  \underbrace{\langle \phi_1 |x| \phi_0 \rangle}_{d_x}  = q E_x d_x, $$

Dipole elements $d_z, d_x$ are approx, Bohr radius $\sim A^{\circ}$. Due to the factor $\frac{\upsilon}{c} \sim 10^{-3}$, $\M_1 \ll \M_1'$. Therefore transition between different atomic orbitals are largely driven by the $x$ term. 

\section{Scattering with $\E \cdot x$ term, the negative sign of amplitude}
Consider Moller scattering with $x$ term
\begin{figure}[htb!]
  \centering
  \includegraphics[scale = .5]{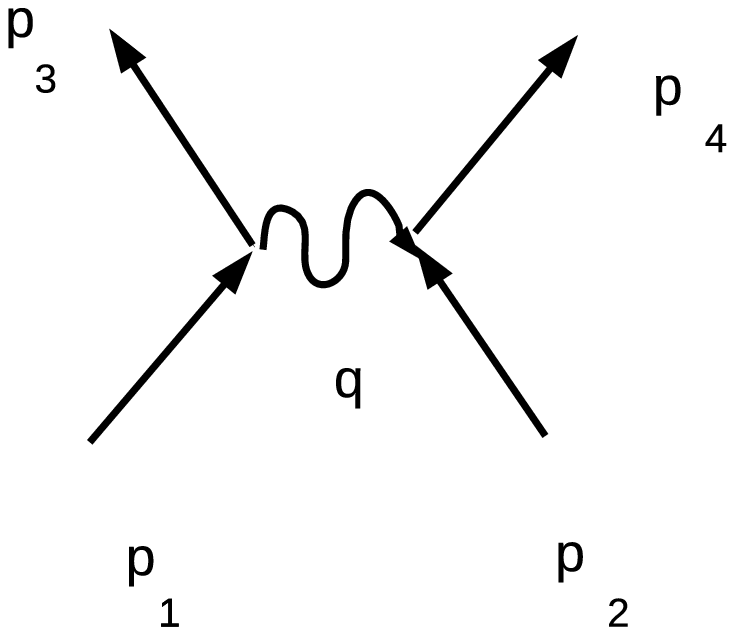}
  \caption{Fig. depicts m\o{}ller scattering. Two electrons with momentum $p_1$ and $p_2$ scatter by exchange of photon to $p_3$ and $p_4$.} \label{fig:moller}
\end{figure}

When electron changes momentum by $q$, in Lorentz gauge, photon of momentum $-q$ is emitted. In $\E \cdot x$ gauge, the emitted photon can be more general with momentum $-q + k$, where $k = n \Delta$ ($\Delta \ \l = 2 \pi$, $l$ length of electron packet) makes angle $\theta$ with $q$, then the amplitude $\M$ of scattering a momentum exchange $q$ is

\begin{equation}
\M_0 = C u_4^\dagger(p_4) u_3(p_2) \ u_2^\dagger(p_3) u_1(p_1) 
\end{equation}

\begin{equation}
C = -\frac{1}{4} \sum_n (\frac{2}{\pi n})^2 \frac{1}{( (q + n \Delta \cos(\theta))^2 + n^2 \Delta^2 \sin^2 \theta )} \int_0^{\pi} \cos^2 \theta d \theta \int_0^{2 \pi} \cos^2 \phi d\phi, 
\end{equation} where $\int_l x \sin(\frac{n 2 \pi}{l} x)= \frac{2}{n \pi}$. Then in limit $\Delta \rightarrow 0$, we get $C = -\frac{1}{|q|^2}$.

We of course have (from $A_z$) the term

\begin{equation}
\M_q = C u_4^\dagger(p_4) \gamma_q u_3(p_2) \ u_2^\dagger(p_3) \gamma^q u_1(p_1) 
\end{equation}

The total amplitude including contribution from $A_x, A_y$ in gauge

\begin{equation}
\M = \frac{1}{4 |q|^2 } \{ u_4^\dagger(p_4) \gamma_\mu u_3(p_2) \ u_2^\dagger(p_3) \gamma^\mu u_1(p_1) - \left ( 3 \M_0 + 5 \M_q    \right ) \}. 
\end{equation}

The first term is the usual Feynman propagator, scaled by $\frac{1}{4}$, but second term is new and gives big contribution to Coulomb potential.

That's it, we have a new propagator. In its full glory it reads

\begin{equation}
\M =  \frac{(e \hbar c)^2}{4 \epsilon_0 V q^2}\{ u_4^\dagger(p_4) \gamma_\mu u_3(p_2) \ u_2^\dagger(p_3) \gamma^\mu u_1(p_1) - \left ( 3 \M_0 + 5 \M_q    \right ) \}. 
\end{equation}


\begin{thebibliography}{99}

\bibitem{peskin} M.E. Peskin and D.V. Schroeder ``An Introduction to Quantum Field Theory'', CRC Press, 1995.

\bibitem{griffiths.0} David J. Griffiths, ``Introduction to Elementary Particles'', John Wiley and Sons (1987).

\bibitem{thomson} Mark Thomson, ``Modern particle Physics'', Cambridge University Press (2013).

\bibitem{Griffiths} David J. Griffiths, ``Introduction to Quantum Mechanics'', Pearson Prentice Hall (2004).   


\end{thebibliography}
\end{document}